\documentclass[a4paper,11pt]{article}
\usepackage{pos}

\newcommand{\muEW}{{\mu_\text{ew}}}
\newcommand{\WcL}[2][{}]{C_{#2}^{#1}}
\newcommand{\muNP}{{\Lambda}}

\newcommand{\wc}[3][{}]{\big[\mathcal{C}_{#2}^{#1}\big]_{#3}}
\newcommand{\DF}{\Delta F}
\newcommand{\OpL}[2][{}]{Q_{#2}^{#1}}
\newcommand{\cH}{\mathcal{H}}
\newcommand{\GeV}{\,\text{GeV}}

\newcommand{\MSbar}{${\overline{\text{MS}}}$}
\newcommand{\bra}[1]{\ensuremath{\langle #1 |}}
\newcommand{\ket}[1]{\ensuremath{| #1 \rangle }}
\newcommand{\oL}[1]{\overline{#1}}
\newcommand{\ord}[1]{\mathcal{O}\left( #1 \right)}
\newcommand{\KKbar}{K^0-\bar K^0}
\newcommand{\DDbar}{D^0-\bar D^0}
\newcommand{\BBbar}{B_{s,d}-\bar B_{s,d}}

\newcommand{\BBbarS}{B_s-\bar B_s}
\newcommand{\vp}{\phi}
\newcommand{\vpj}{\vp^\dagger i \overleftrightarrow{\mathcal{D}}_{\!\!\!\mu} \vp}
\newcommand{\Op}[2][{}]{\mathcal{O}_{#2}^{#1}}

\title{SMEFT interpretation of $\Delta$F = 2 transitions}
\author*[a]{Jason~Aebischer}

\affiliation[a]{Physik-Institut, Universit\"at Z\"urich, CH-8057 Z\"urich, Switzerland}

\emailAdd{jason.aebischer@physik.uzh.ch}

\abstract{A model-independent anatomy of $\Delta F= 2$  transitions in the context of the Weak Effective Theory (WET) below the electroweak scale (EW) and the Standard Model Effective Field Theory (SMEFT) above the EW scale is discussed. Two master formulae for the BSM contribution of the mixing amplitude $M_{12}$, in terms of Wilson coefficients are presented.
The coefficients entering these formulae contain all the information below the EW scale and the NP scale $\Lambda$, respectively. The renormalization group evolution from the top-quark Yukawa coupling has the largest impact on the result. The obtained expressions depend on whether the down-basis or the up-basis for SMEFT operators is considered. }

\FullConference{%
  11th International Workshop on the CKM Unitarity Triangle (CKM2021)\\
  22-26 November 2021\\
  The University of Melbourne, Australia
}

\begin{document}
\maketitle

\section{Introduction}
Processes involving a flavour change by two units, so called $\Delta F=2$ processes, are one of the most powerful means to test New Physics (NP) effects that go beyond the Standard Model (SM). They arise in the neutral meson mixing processes $D^0-\bar{D}^0,\,\, K^0-\bar{K}^0,\,\, B_{s,d}^0-\bar{B}_{s,d}^0$.
These processes can probe NP scales which exceed the reach of the LHC at CERN by several orders of magnitude and are therefore of great interest in the search for physics beyond the SM.

The mixing of the neutral meson decays is described by the Schr\"odinger equation

\begin{equation}
  i \frac{d}{dt}|\psi\rangle =\hat H|\psi\rangle\,,
\end{equation}
for the two-component state $|\psi\rangle$ describing a neutral meson $M^0$ and its anti-particle $\oL{M^0}$

\begin{equation}
  |\psi\rangle = \begin{pmatrix}|M^0\rangle\\ |\oL{M^0}\rangle\end{pmatrix}\,,
\end{equation}
governed by the Hamiltonian

\begin{equation}
  \hat H = \hat M -\frac{i}{2}\hat \Gamma = \begin{pmatrix}M -\frac{i}{2} \Gamma & M_{12} -\frac{i}{2} \Gamma_{12} \\ M_{12}^* -\frac{i}{2} \Gamma_{12}^* & M -\frac{i}{2} \Gamma\end{pmatrix}\,,
\end{equation}
where $CPT$ invariance has been imposed. The matrices $\hat M$ and $\hat \Gamma$ are hermitian and describe off-shell and on-shell transitions between the two neutral mesons states, respectively. In the following we will focus on the virtual contributions which generate neutral meson mixing, characterized by the off-diagonal matrix element $M_{12}$. Physical observables involving meson mixing are expressed as a function of $M_{12}$, like the mixing parameter $x_{12}=2|M_{12}|/\Gamma$ in the $D^0-\bar{D}^0$ system, with the averaged decay rate $\Gamma$. The quantity $M_{12}$ can be split into a SM and a beyond the SM (BSM) piece as follows:

\begin{equation}
  M_{12}=\big[M_{12}\big]_{\text{SM}}+\big[M_{12}\big]_{\text{BSM}}\,.
\end{equation}

Following the discussion in \cite{Aebischer:2020dsw}, we will focus on the BSM part of the matrix element $M_{12}$, which for a neutral meson $M^0$ is given by
\begin{align}
  \label{eq:def-M12BSM}
  \big[M_{12}^{ij}\big]_\text{BSM} &
  = \frac{1}{2 m_{M^0}} \bra{{M}^0} \cH_{\DF=2}^{ij} \ket{\oL{M^0}} (\mu)
    {\, + \, \ord{\text{dim-8}}} \,,
\end{align}
where $ m_{M^0}$ denotes the mass of the neutral meson and $\cH_{\DF=2}^{ij}$ is the effective Hamiltonian governing $\DF=2$ transitions of the form $j\to i$. In our convention the meson mixing processes $D^0-\bar{D}^0,\,\, K^0-\bar{K}^0,\,\, B_{s,d}^0-\bar{B}_{s,d}^0$ are described by the flavour indices $ij = cu, sd, sb, db$. For simplicity we neglect contributions from dimension-eight operators in eq.~\eqref{eq:def-M12BSM}.

In the following sections we will derive a master formula similar as in the case for $\varepsilon'/\varepsilon$ \cite{Aebischer:2018quc,Aebischer:2018csl,Aebischer:2020jto,Aebischer:2021hws}, the $(g-2)_\mu$ \cite{Aebischer:2021uvt,Aebischer:2021pei} and $\DF=1$ transitions \cite{Aebischer:2021raf}, but now for the matrix element $\big[M_{12}^{ij}\big]_\text{BSM}$; first in the Weak Effective Theory (WET) \cite{Aebischer:2017gaw} valid below the electroweak (EW) scale, discussed in Section~\ref{sec:WET}, and secondly in the SM Effective Theory (SMEFT) above the EW scale in Section~\ref{sec:SMEFT}. The findings will be summarized in Section~\ref{sec:summary}.

\section{WET master formula}\label{sec:WET}
The WET is the effective theory valid below the EW scale which includes all SM particles except the $W,Z,t,H$. In this theory the interaction Hamiltonian governing $\DF=2$ processes reads:

  \begin{align}
    \label{eq:DF2-Heff}
    \cH_{\DF=2}^{ij} &
    =   \sum_a \WcL[ij]{a}\, \OpL[ij]{a}
    + \text{h.c.} \,,
  \end{align}

\noindent
where the sum runs over local operators $\OpL[ij]{a}$, which are weighted by the corresponding Wilson coefficients $\WcL[ij]{a}$. In the case of $D^0-\bar D^0$ the operators take the form:

  \begin{equation*}
    \label{eq:BMU-basis}
  \begin{aligned}
    \OpL[cu]{\text{VLL}} &
    = [\bar{c} \gamma_\mu P_L u][\bar{c} \gamma^\mu P_L u]\,, &
  \\[0.2cm]
    \OpL[cu]{\text{LR},1} &
    = [\bar{c} \gamma_\mu P_L u][\bar{c} \gamma^\mu P_R u]\,, &
    \OpL[cu]{\text{LR},2} &
    = [\bar{c} P_L u][\bar{c} P_R u]\,,
  \\[0.2cm]
    \OpL[cu]{\text{SLL},1} &
    = [\bar{c} P_L u][\bar{c} P_L u]\,, &
    \OpL[cu]{\text{SLL},2} &
    = -[\bar{c} \sigma_{\mu\nu} P_L u][\bar{c} \sigma^{\mu\nu} P_L u] \,,
  \end{aligned}
  \end{equation*}

\noindent
and similar for the other neutral meson mixings with obvious flavour replacements. These operators can then be used to compute the BSM contribution to the $M_{12}$ element via eq.~\eqref{eq:def-M12BSM}. For this purpose the matrix elements of the operators have to be computed, which we take from Lattice QCD (LQCD):
\begin{itemize}
  \item
    For the case of $\DDbar$ mixing, FNAL/MILC calculated bag factors for $N_f = 2+1$~\cite{Bazavov:2017weg} for the full set of
    $\Delta C = 2$ operators, which we use in our calculations.
\item
  In the case of $\KKbar$ mixing we use the FLAG average for $N_f = 2+1$ \cite{Boyle:2017ssm,Aoki:2019cca}.
\item
  For the case of $\BBbar$ mixing the full set of matrix elements has
  been calculated with LQCD methods for
  $N_f = 2 + 1$ by FNAL/MILC \cite{Bazavov:2016nty} and for $N_f = 2+ 1 +1$
  by HPQCD \cite{Dowdall:2019bea}. We use the averages of the
  bag factors from HPQCD and FNAL/MILC, as given in \cite{Dowdall:2019bea}.
\end{itemize}

As indicated by the $\mu$-dependence in eq.~\eqref{eq:def-M12BSM}, the matrix elements from LQCD are evaluated at a certain renormalization scale, which is different for each of the meson mixing processes. In order to cancel this scale dependence, the Wilson coefficients of the effective Hamiltonian in eq.~\eqref{eq:DF2-Heff} have to be evaluated at the same scale, using renormalization group equations (RGEs). For this purpose we use the one-loop and two-loop RGEs \cite{Buras:2000if} for the Wilson coefficients in question. The one-loop running is performed, using the Python package \texttt{wilson} \cite{Aebischer:2018bkb} and adopting the \texttt{WCxf} convention \cite{Aebischer:2017ugx} for the Wilson coefficients.

The consideration of two-loop RGE effects introduces further scheme-dependence in the calculation in the form of evanescent operators. The careful treatment of these effects, using the shift method outlined in \cite{Aebischer:2022tvz} is discussed in detail in Ref.~\cite{Aebischer:2020dsw}.

\begin{figure}[t]
  \begin{center}
    \includegraphics[width=0.45\textwidth]{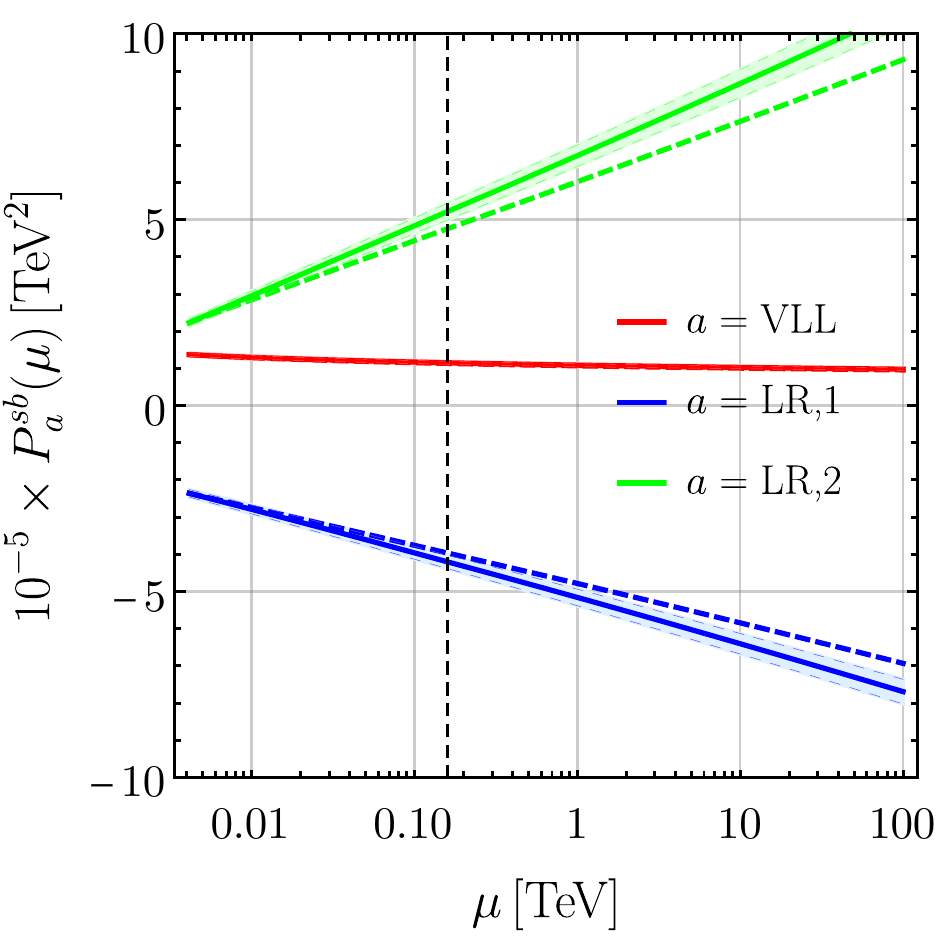}
  \end{center}
    \caption{\small The QCD RG evolution at NLO [solid] versus LO [dashed]
      for some of the coefficients $P_a^{ij}(\mu)$ for
      $\BBbarS$ mixing. The coloured band around the NLO
      results shows the hadronic uncertainties from the matrix elements.
      The vertical dashed line indicates $\muEW = 160\GeV$.
    }
    \label{fig:qcd-1-2}
\end{figure}

Having now all the relevant ingredients at hand, the master formula for $\big[M_{12}^{ij}\big]_\text{BSM}$ at the EW scale $\muEW$ can be written as

\begin{equation}
  \label{eq:master-M12BSM}
  2\big[M_{12}^{ij}\big]_\text{BSM}
  = (\Delta M_{ij})_\text{exp}
    \sum_{a}^{\phantom{x}} P_a^{ij}(\muEW) \, \WcL[ij]{a}(\muEW)\,,
\end{equation}
where the expression is normalized to the experimental mass difference of the mesons, $(\Delta M)_{\text{exp}}$, and where the $P_a$ factors contain the hadronic matrix elements as well as the RG evolution of the Wilson coefficients from the EW scale down to the hadronic scale. The $P_a$ factors evaluated at $\muEW$ are shown for each sector in Tab.~\ref{tab:LEFT-Pa}, and are given for other bases in Ref.~\cite{Aebischer:2020dsw}.
The QCD evolution of several $P_a^{sb}$ factors is shown in Fig.~\ref{fig:qcd-1-2}. It is largest ($\sim$10\%) for the LR operators and negligible for the VLL one. Similar conclusions can be drawn for the other sectors.

\begin{table}[t]
\centering
\renewcommand{\arraystretch}{1.2}
\begin{tabular}{|c|ccccc|c|}
\hline
     & $P_\text{VLL}$ & $P_{\text{SLL},1}$ & $P_{\text{SLL},2}$ & $P_{\text{LR},1}$ & $P_{\text{LR},2}$ \\
\hline
  $K^0$ &       0.102(2) &     -4.32(16) &     -7.93(37) &     -8.55(28) &     14.14(82)  \\
  $D^0$ &        0.56(4) &     -2.20(11) &     -4.04(28) &     -4.23(22) &      6.18(44)  \\
  $B_d$ &       2.67(10) &     -4.99(28) &     -9.05(68) &    -10.29(54) &     12.75(50)  \\
  $B_s$ &       1.15(4)  &     -2.24(13) &     -4.08(26) &     -4.20(18) &      5.22(21)  \\
\hline
\end{tabular}
\renewcommand{\arraystretch}{1.0}
\caption{\small
  \label{tab:LEFT-Pa}
  The values of the coefficients $P_a^{ij}(\muEW)$ entering the WET master formula
  in Eq.~\eqref{eq:master-M12BSM} at $\muEW = 160 \GeV$
  using as input the \MSbar{}-NDR matrix elements at the low-energy scale.
  The shown uncertainties are due to the matrix elements or the bag factors and their corresponding
  chiral enhancement factors.
}
\end{table}

\section{SMEFT master formula}\label{sec:SMEFT}
In this section we discuss the analogous master formula as in eq.~\eqref{eq:master-M12BSM}, but in the context of SMEFT degrees of freedom. The SMEFT is the effective theory above the EW scale, incorporating the full SM content \cite{Grzadkowski:2010es}. The part of the SMEFT master formula below the EW scale is identical to the one in the previous section. The only difference is the addition of the tree-level \cite{Jenkins:2017jig} and one-loop matching \cite{Aebischer:2015fzz,Dekens:2019ept} from SMEFT onto WET, as well as the SMEFT running above the EW scale \cite{Jenkins:2013zja,Jenkins:2013wua,Alonso:2013hga}, which again has been performed using \texttt{wilson}. The SMEFT master formula is then given by

\begin{equation}\label{eq:SMEFTMF}
  2 \big [M_{12}^{ij} \big]_\text{BSM}
  = (\Delta M_{ij})_\text{exp} \sum_{a} \,
    P_a^{ij}(\muNP) \; \wc{a}{ij}(\muNP)\,,
\end{equation}
where now the sum runs over the SMEFT Wilson coefficients. The values of the SMEFT $P_a$ factors are given in Ref.~\cite{Aebischer:2020dsw}. Their form depends on the basis chosen for the SMEFT operators. In particular, the Yukawa running effects and the resulting back-rotation of the Wilson coefficients \cite{Aebischer:2020lsx} has to be taken into account, when changing from one basis to another.

\begin{figure}
  \begin{center}
    \includegraphics[width=0.6\textwidth]{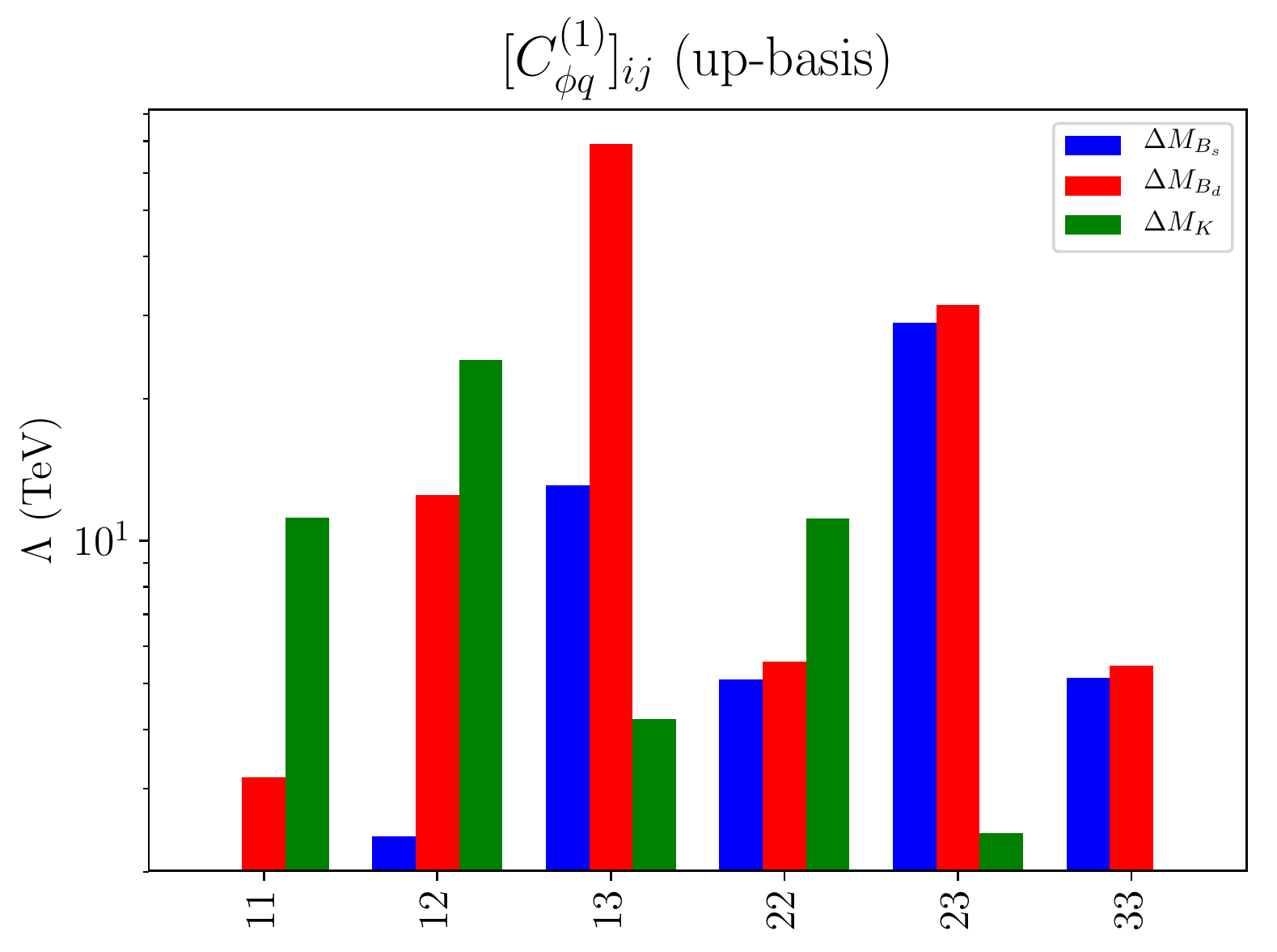}
  \end{center}
  \caption{\small
    The maximal NP scale $\muNP$ for a $10\%$ effect in $2\big[M_{12}^{ij}\big]_\text{BSM}/
    (\Delta M_{ij})_\text{exp}$ for $B_s$ (blue), $B_d$ (red) and $K^0$
    (green), respectively.
  }\label{fig:bounds}
\end{figure}

Applying eq.~\eqref{eq:SMEFTMF} to the Wilson coefficients of the operators

\begin{equation}
  [\Op[(1)]{\vp q}]_{ij} = (\vpj)(\bar q^i \gamma^\mu q^j)\,,
\end{equation}
leads to the bounds for the NP scale $\Lambda$ shown in Fig.~\ref{fig:bounds} for the different sectors. The bounds range from several to about a hundred TeV, depending on the process and flavour structure of the operator.

\section{Summary}\label{sec:summary}
We have presented a master formula for the Wilson coefficients describing $\DF=2$ effects
in the WET below the EW scale, as well as in the SMEFT above the EW scale. Two-loop running below the EW scale has been taken account, as well as one-loop matching and running contributions in the SMEFT. The formulae are completely general and allow to compute the contribution of any model to neutral meson mixing, provided it has been matched onto the SMEFT or WET.

\small
\bibliographystyle{JHEP}
\bibliography{Bookallrefs}

\end{document}